
\catcode`\@=11

\font\fourteenrm=cmr10 scaled\magstep2
\font\twelverm=cmr10 scaled\magstep1
\font\ninerm=cmr9	     \font\sixrm=cmr6
\font\seventeenbf=cmbx10 scaled\magstep3
\font\fourteenbf=cmbx10 scaled\magstep2
\font\twelvebf=cmbx10 scaled\magstep1
\font\ninebf=cmbx9	      \font\sixbf=cmbx6
\font\seventeeni=cmmi10 scaled\magstep3	    \skewchar\seventeeni='177
\font\fourteeni=cmmi10 scaled\magstep2	    \skewchar\fourteeni='177
\font\twelvei=cmmi10 scaled\magstep1	    \skewchar\twelvei='177
\font\ninei=cmmi9			    \skewchar\ninei='177
\font\sixi=cmmi6			    \skewchar\sixi='177
\font\seventeensy=cmsy10 scaled\magstep3    \skewchar\seventeensy='60
\font\fourteensy=cmsy10 scaled\magstep2	    \skewchar\fourteensy='60
\font\twelvesy=cmsy10 scaled\magstep1	    \skewchar\twelvesy='60
\font\ninesy=cmsy9			    \skewchar\ninesy='60
\font\sixsy=cmsy6			    \skewchar\sixsy='60

\font\fourteenex=cmex10 scaled\magstep2
\font\twelveex=cmex10 scaled\magstep1

\font\fourteensl=cmsl10 scaled\magstep2
\font\twelvesl=cmsl10 scaled\magstep1
\font\ninesl=cmsl9

\font\fourteenit=cmti10 scaled\magstep2
\font\twelveit=cmti10 scaled\magstep1
\font\twelvett=cmtt10 scaled\magstep1
\font\twelvecp=cmcsc10 scaled\magstep1
\font\tencp=cmcsc10
\newfam\cpfam
\newcount\f@ntkey	     \f@ntkey=0
\def\samef@nt{\relax \ifcase\f@ntkey \rm \or\oldstyle \or\or
	 \or\it \or\sl \or\bf \or\tt \or\caps \fi }
\def\fourteenpoint{\relax
   \textfont0=\fourteenrm	    \scriptfont0=\tenrm
   \scriptscriptfont0=\sevenrm
   \def\rm{\fam0 \fourteenrm \f@ntkey=0 }%
   \textfont1=\fourteeni	    \scriptfont1=\teni
   \scriptscriptfont1=\seveni
   \def\oldstyle{\fam1 \fourteeni\f@ntkey=1 }%
   \textfont2=\fourteensy	    \scriptfont2=\tensy
   \scriptscriptfont2=\sevensy
   \textfont3=\fourteenex     \scriptfont3=\fourteenex
   \scriptscriptfont3=\fourteenex
   \def\it{\fam\itfam \fourteenit\f@ntkey=4 }\textfont\itfam=\fourteenit
   \def\sl{\fam\slfam \fourteensl\f@ntkey=5 }\textfont\slfam=\fourteensl
   \scriptfont\slfam=\tensl
   \def\bf{\fam\bffam \fourteenbf\f@ntkey=6 }\textfont\bffam=\fourteenbf
   \scriptfont\bffam=\tenbf	 \scriptscriptfont\bffam=\sevenbf
   \def\tt{\fam\ttfam \twelvett \f@ntkey=7 }\textfont\ttfam=\twelvett
   \h@big=11.9\p@ \h@Big=16.1\p@ \h@bigg=20.3\p@ \h@Bigg=24.5\p@
   \def\caps{\fam\cpfam \twelvecp \f@ntkey=8 }\textfont\cpfam=\twelvecp
   \setbox\strutbox=\hbox{\vrule height 12pt depth 5pt width\z@}%
   \samef@nt}
\def\twelvepoint{\relax
   \textfont0=\twelverm	  \scriptfont0=\ninerm
   \scriptscriptfont0=\sevenrm
   \def\rm{\fam0 \twelverm \f@ntkey=0 }%
   \textfont1=\twelvei		  \scriptfont1=\ninei
   \scriptscriptfont1=\seveni
   \def\oldstyle{\fam1 \twelvei\f@ntkey=1 }%
   \textfont2=\twelvesy	  \scriptfont2=\ninesy
   \scriptscriptfont2=\sevensy
   \textfont3=\twelveex	  \scriptfont3=\twelveex
   \scriptscriptfont3=\twelveex
   \def\it{\fam\itfam \twelveit \f@ntkey=4 }\textfont\itfam=\twelveit
   \def\sl{\fam\slfam \twelvesl \f@ntkey=5 }\textfont\slfam=\twelvesl
   \scriptfont\slfam=\ninesl
   \def\bf{\fam\bffam \twelvebf \f@ntkey=6 }\textfont\bffam=\twelvebf
   \scriptfont\bffam=\ninebf	  \scriptscriptfont\bffam=\sevenbf
   \def\tt{\fam\ttfam \twelvett \f@ntkey=7 }\textfont\ttfam=\twelvett
   \h@big=10.2\p@ \h@Big=13.8\p@ \h@bigg=17.4\p@ \h@Bigg=21.0\p@
   \def\caps{\fam\cpfam \twelvecp \f@ntkey=8 }\textfont\cpfam=\twelvecp
   \setbox\strutbox=\hbox{\vrule height 10pt depth 4pt width\z@}%
   \samef@nt}
\def\tenpoint{\relax
   \textfont0=\tenrm	       \scriptfont0=\sevenrm
   \scriptscriptfont0=\fiverm
   \def\rm{\fam0 \tenrm \f@ntkey=0 }%
   \textfont1=\teni	       \scriptfont1=\seveni
   \scriptscriptfont1=\fivei
   \def\oldstyle{\fam1 \teni \f@ntkey=1 }%
   \textfont2=\tensy	       \scriptfont2=\sevensy
   \scriptscriptfont2=\fivesy
   \textfont3=\tenex	       \scriptfont3=\tenex
   \scriptscriptfont3=\tenex
   \def\it{\fam\itfam \tenit \f@ntkey=4 }\textfont\itfam=\tenit
   \def\sl{\fam\slfam \tensl \f@ntkey=5 }\textfont\slfam=\tensl
   \def\bf{\fam\bffam \tenbf \f@ntkey=6 }\textfont\bffam=\tenbf
   \scriptfont\bffam=\sevenbf	   \scriptscriptfont\bffam=\fivebf
   \def\tt{\fam\ttfam \tentt \f@ntkey=7 }\textfont\ttfam=\tentt
   \def\caps{\fam\cpfam \tencp \f@ntkey=8 }\textfont\cpfam=\tencp
   \h@big=8.5\p@ \h@Big=11.5\p@ \h@bigg=14.5\p@ \h@Bigg=17.5\p@
   \setbox\strutbox=\hbox{\vrule height 8.5pt depth 3.5pt width\z@}%
   \samef@nt}
\newdimen\h@big  \h@big=8.5\p@
\newdimen\h@Big  \h@Big=11.5\p@
\newdimen\h@bigg  \h@bigg=14.5\p@
\newdimen\h@Bigg  \h@Bigg=17.5\p@
\def\big#1{{\hbox{$\left#1\vbox to\h@big{}\right.\n@space$}}}
\def\Big#1{{\hbox{$\left#1\vbox to\h@Big{}\right.\n@space$}}}
\def\bigg#1{{\hbox{$\left#1\vbox to\h@bigg{}\right.\n@space$}}}
\def\Bigg#1{{\hbox{$\left#1\vbox to\h@Bigg{}\right.\n@space$}}}
\normalbaselineskip = 20pt plus 0.2pt minus 0.1pt
\normallineskip = 1.5pt plus 0.1pt minus 0.1pt
\normallineskiplimit = 1.5pt
\newskip\normaldisplayskip
\normaldisplayskip = 20pt plus 5pt minus 10pt
\newskip\normaldispshortskip
\normaldispshortskip = 6pt plus 5pt
\newskip\normalparskip
\normalparskip = 6pt plus 2pt minus 1pt
\newskip\skipregister
\skipregister = 5pt plus 2pt minus 1.5pt
\newif\ifsingl@	   \newif\ifdoubl@
\newif\iftwelv@	   \twelv@true
\def\singlespace{\singl@true\doubl@false\spaces@t}
\def\doublespace{\singl@false\doubl@true\spaces@t}
\def\normalspace{\singl@false\doubl@false\spaces@t}
\def\Tenpoint{\tenpoint\twelv@false\spaces@t}
\def\Twelvepoint{\twelvepoint\twelv@true\spaces@t}
\def\spaces@t{\relax
   \iftwelv@\ifsingl@\subspaces@t3:4;\else\subspaces@t1:1;\fi
   \else\ifsingl@\subspaces@t3:5;\else\subspaces@t4:5;\fi\fi
   \ifdoubl@\multiply\baselineskip by 5 \divide\baselineskip by 4 \fi}
\def\subspaces@t#1:#2;{\baselineskip=\normalbaselineskip
   \multiply\baselineskip by #1\divide\baselineskip by #2%
   \lineskip = \normallineskip
   \multiply\lineskip by #1\divide\lineskip by #2%
   \lineskiplimit = \normallineskiplimit
   \multiply\lineskiplimit by #1\divide\lineskiplimit by #2%
   \parskip = \normalparskip
   \multiply\parskip by #1\divide\parskip by #2%
   \abovedisplayskip = \normaldisplayskip
   \multiply\abovedisplayskip by #1\divide\abovedisplayskip by #2%
   \belowdisplayskip = \abovedisplayskip
   \abovedisplayshortskip = \normaldispshortskip
   \multiply\abovedisplayshortskip by #1%
   \divide\abovedisplayshortskip by #2%
   \belowdisplayshortskip = \abovedisplayshortskip
   \advance\belowdisplayshortskip by \belowdisplayskip
   \divide\belowdisplayshortskip by 2
   \smallskipamount = \skipregister
   \multiply\smallskipamount by #1\divide\smallskipamount by #2%
   \medskipamount = \smallskipamount \multiply\medskipamount by 2
   \bigskipamount = \smallskipamount \multiply\bigskipamount by 4 }
\def\normalbaselines{ \baselineskip=\normalbaselineskip%
   \lineskip=\normallineskip \lineskiplimit=\normallineskip%
   \iftwelv@\else \multiply\baselineskip by 4 \divide\baselineskip by 5%
   \multiply\lineskiplimit by 4 \divide\lineskiplimit by 5%
   \multiply\lineskip by 4 \divide\lineskip by 5 \fi }
\Twelvepoint  
\interlinepenalty=50
\interfootnotelinepenalty=5000
\predisplaypenalty=9000
\postdisplaypenalty=500
\hfuzz=1pt
\vfuzz=0.2pt
%
%
%
\def\pagecontents{%
   \ifvoid\topins\else\unvbox\topins\vskip\skip\topins\fi
   \dimen@ = \dp255 \unvbox255
   \ifvoid\footins\else\vskip\skip\footins\footrule\unvbox\footins\fi
   \ifr@ggedbottom \kern-\dimen@ \vfil \fi }
\def\makeheadline{\vbox to 0pt{ \skip@=\topskip
   \advance\skip@ by -12pt \advance\skip@ by -2\normalbaselineskip
   \vskip\skip@ \line{\vbox to 12pt{}\the\headline} \vss
   }\nointerlineskip}
\def\makefootline{\baselineskip = 1.5\normalbaselineskip
   \line{\the\footline}}
\newif\iffrontpage
\newif\ifp@genum
\def\nopagenumbers{\p@genumfalse}
\def\pagenumbers{\p@genumtrue}
\pagenumbers
\newtoks\date
\newtoks\Month
\footline={\hss\iffrontpage\else\ifp@genum\tenrm\folio\hss\fi\fi}
\headline={\iffinal\hfil\else\tenrm DRAFT\hfil\the\date\fi}
\def\monthname{\relax\ifcase\month 0/\or January\or February\or
   March\or April\or May\or June\or July\or August\or September\or
   October\or November\or December\else\number\month/\fi}
\date={\monthname\ \number\day, \number\year}
\Month={\monthname\ \number\year}
\countdef\pagenumber=1  \pagenumber=1
\def\advancepageno{\global\advance\pageno by 1
   \ifnum\pagenumber<0 \global\advance\pagenumber by -1
   \else\global\advance\pagenumber by 1 \fi \global\frontpagefalse }
\def\folio{\ifnum\pagenumber<0 \romannumeral-\pagenumber
   \else \number\pagenumber \fi }
\def\footrule{\dimen@=\prevdepth\nointerlineskip
   \vbox to 0pt{\vskip -0.25\baselineskip \hrule width 0.35\hsize \vss}
   \prevdepth=\dimen@ }
\newtoks\foottokens
\foottokens={\Tenpoint\singlespace}
\newdimen\footindent
\footindent=24pt
\def\vfootnote#1{\insert\footins\bgroup  \the\foottokens
   \interlinepenalty=\interfootnotelinepenalty \floatingpenalty=20000
   \splittopskip=\ht\strutbox \boxmaxdepth=\dp\strutbox
   \leftskip=\footindent \rightskip=\z@skip
   \parindent=0.5\footindent \parfillskip=0pt plus 1fil
   \spaceskip=\z@skip \xspaceskip=\z@skip
   \Textindent{$ #1 $}\footstrut\futurelet\next\fo@t}
\def\Textindent#1{\noindent\llap{#1\enspace}\ignorespaces}
\def\footnote#1{\attach{#1}\vfootnote{#1}}

\let\footsymbol=\star
\newcount\lastf@@t	     \lastf@@t=-1
\newcount\footsymbolcount    \footsymbolcount=0
\newif\ifPhysRev
\def\footsymbolgen{\relax \ifPhysRev \iffrontpage \NPsymbolgen\else
   \PRsymbolgen\fi \else \NPsymbolgen\fi
   \global\lastf@@t=\pageno \footsymbol }
\def\NPsymbolgen{\ifnum\footsymbolcount<0 \global\footsymbolcount=0\fi
   {\iffrontpage \else \advance\lastf@@t by 1 \fi
   \ifnum\lastf@@t<\pageno \global\footsymbolcount=0
   \else \global\advance\footsymbolcount by 1 \fi }
   \ifcase\footsymbolcount \fd@f\star\or \fd@f\dagger\or \fd@f\ast\or
   \fd@f\ddagger\or \fd@f\natural\or \fd@f\diamond\or \fd@f\bullet\or
   \fd@f\nabla\else \fd@f\dagger\global\footsymbolcount=0 \fi }
\def\fd@f#1{\xdef\footsymbol{#1}}
\def\PRsymbolgen{\ifnum\footsymbolcount>0 \global\footsymbolcount=0\fi
   \global\advance\footsymbolcount by -1
   \xdef\footsymbol{\sharp\number-\footsymbolcount} }
\def\space@ver#1{\let\@sf=\empty \ifmmode #1\else \ifhmode
   \edef\@sf{\spacefactor=\the\spacefactor}\unskip${}#1$\relax\fi\fi}
\def\attach#1{\space@ver{\strut^{\mkern 2mu #1} }\@sf\ }
%
%
%
\newcount\chapternumber	     \chapternumber=0
\newcount\sectionnumber	     \sectionnumber=0
\newcount\equanumber	     \equanumber=0
\let\chapterlabel=\relax
\newtoks\chapterstyle	     \chapterstyle={\Number}
\newskip\chapterskip	     \chapterskip=\bigskipamount
\newskip\sectionskip	     \sectionskip=\medskipamount
\newskip\headskip	     \headskip=8pt plus 3pt minus 3pt
\newdimen\chapterminspace    \chapterminspace=15pc
\newdimen\sectionminspace    \sectionminspace=10pc
\newdimen\referenceminspace  \referenceminspace=25pc
\def\chapterreset{\global\advance\chapternumber by 1
   \ifnum\equanumber<0 \else\global\equanumber=0\fi
   \sectionnumber=0 \makel@bel}
\def\makel@bel{\xdef\chapterlabel{%
   \the\chapterstyle{\the\chapternumber}.}}
\def\sectionlabel{\number\sectionnumber \quad }
\def\alphabetic#1{\count255='140 \advance\count255 by #1\char\count255}
\def\Alphabetic#1{\count255='100 \advance\count255 by #1\char\count255}
\def\Roman#1{\uppercase\expandafter{\romannumeral #1}}
\def\roman#1{\romannumeral #1}
\def\Number#1{\number #1}
\def\unnumberedchapters{\let\makel@bel=\relax \let\chapterlabel=\relax
\let\sectionlabel=\relax \equanumber=-1 }
\def\titlestyle#1{\par\begingroup \interlinepenalty=9999
   \leftskip=0.02\hsize plus 0.23\hsize minus 0.02\hsize
   \rightskip=\leftskip \parfillskip=0pt
   \hyphenpenalty=9000 \exhyphenpenalty=9000
   \tolerance=9999 \pretolerance=9000
   \spaceskip=0.333em \xspaceskip=0.5em
   \iftwelv@\fourteenpoint\else\twelvepoint\fi
   \noindent #1\par\endgroup }
\def\spacecheck#1{\dimen@=\pagegoal\advance\dimen@ by -\pagetotal
   \ifdim\dimen@<#1 \ifdim\dimen@>0pt \vfil\break \fi\fi}
\def\chapter#1{\par \penalty-300 \vskip\chapterskip
   \spacecheck\chapterminspace
   \chapterreset \titlestyle{\chapterlabel \ #1}
   \nobreak\vskip\headskip \penalty 30000
   \wlog{\string\chapter\ \chapterlabel} }

\def\section#1{\par \ifnum\the\lastpenalty=30000\else
   \penalty-200\vskip\sectionskip \spacecheck\sectionminspace\fi
   \wlog{\string\section\ \chapterlabel \the\sectionnumber}
   \global\advance\sectionnumber by 1  \noindent
   {\caps\enspace\chapterlabel \sectionlabel #1}\par
   \nobreak\vskip\headskip \penalty 30000 }
\def\subsection#1{\par
   \ifnum\the\lastpenalty=30000\else \penalty-100\smallskip \fi
   \noindent\undertext{#1}\enspace \vadjust{\penalty5000}}

\def\undertext#1{\vtop{\hbox{#1}\kern 1pt \hrule}}
%

%
\def\APPENDIX#1#2{\par\penalty-300\vskip\chapterskip
   \spacecheck\chapterminspace \chapterreset \xdef\chapterlabel{#1}
   \titlestyle{APPENDIX #2} \nobreak\vskip\headskip \penalty 30000
   \wlog{\string\Appendix\ \chapterlabel} }
\def\Appendix#1{\APPENDIX{#1}{#1}}
\def\appendix{\APPENDIX{A}{}}
%
%
%
\newif\iffinal \finaltrue
\def\showeqname#1{\iffinal\else\hbox to 0pt{\tentt\kern2mm\string#1\hss}\fi}
\def\showEqname#1{\iffinal\else \hskip 0pt plus 1fill
 \hbox to 0pt{\tentt\kern2mm\string#1\hss}\hskip 0pt plus -1fill\fi}
\def\eqnamedef#1{\relax \ifnum\equanumber<0
   \xdef#1{{\noexpand\rm(\number-\equanumber)}}\global\advance\equanumber by -1
   \else \global\advance\equanumber by 1
   \xdef#1{{\noexpand\rm(\chapterlabel \number\equanumber)}}\fi}
\def\eqnamenewdef#1#2{\relax \ifnum\equanumber<0
   \xdef#1{{\noexpand\rm(\number-\equanumber#2)}}\global\advance\equanumber
   by -1 \else \global\advance\equanumber by 1
   \xdef#1{{\noexpand\rm(\chapterlabel \number\equanumber#2)}}\fi}
\def\eqnameolddef#1#2{\relax \ifnum\equanumber<0
   \global\advance\equanumber by 1
   \xdef#1{{\noexpand\rm(\number-\equanumber#2)}}\global\advance\equanumber
   by -1 \else \xdef#1{{\noexpand\rm(\chapterlabel \number\equanumber#2)}}\fi}
\def\eqname#1{\eqnamedef{#1}#1}
\def\eqnamenew#1#2{\eqnamenewdef{#1}{#2}#1}
\def\eqnameold#1#2{\eqnameolddef{#1}{#2}#1}
\def\eq{\eqname\lasteq}
\def\eqa{\eqnamenew\lasteq a}
\def\eqb{\eqnameold\lasteq b}
\def\eqc{\eqnameold\lasteq c}
\def\eqd{\eqnameold\lasteq d}
\def\eqnew#1{\eqnamenew\lasteq{#1}}
\def\eqold#1{\eqnameold\lasteq{#1}}
\def\eq@@{\ifinner\let\eqn@=\relax\else\let\eqn@=\eqno\fi\eqn@}
\def\Eq{\eq@@\eq}
\def\Eqnew#1{\eq@@\eqnew{#1}}
\def\Eqold#1{\eq@@\eqold{#1}}
\def\Eqa{\eq@@\eqa}
\def\Eqb{\eq@@\eqb}
\def\Eqc{\eq@@\eqc}
\def\Eqd{\eq@@\eqd}
\def\Eqn#1{\eq@@\eqname{#1}\showeqname{#1}}
\def\Eqnnew#1#2{\eq@@\eqnamenew{#2}{#1}\showeqname{#1}}
\def\Eqnold#1#2{\eq@@\eqnameold{#2}{#1}\showeqname{#1}}
\def\Eqna#1{\eq@@\eqnamenew{#1}a\showeqname{#1}}
\def\Eqnb#1{\eq@@\eqnameold{#1}b\showeqname{#1}}
\def\Eqnc#1{\eq@@\eqnameold{#1}c\showeqname{#1}}
\def\Eqnd#1{\eq@@\eqnameold{#1}d\showeqname{#1}}

\def\sequentialequations{\equanumber=-1}
%
%
\def\GENITEM#1;#2{\par \hangafter=0 \hangindent=#1
   \Textindent{$ #2 $}\ignorespaces}
\outer\def\newitem#1=#2;{\gdef#1{\GENITEM #2;}}
\newdimen\itemsize		  \itemsize=30pt
\newitem\item=1\itemsize;
\newitem\sitem=1.75\itemsize;	  
\newitem\ssitem=2.5\itemsize;	  
\outer\def\newlist#1=#2&#3&#4;{\toks0={#2}\toks1={#3}%
   \count255=\escapechar \escapechar=-1
   \alloc@0\list\countdef\insc@unt\listcount	 \listcount=0
   \edef#1{\par
      \countdef\listcount=\the\allocationnumber
      \advance\listcount by 1
      \hangafter=0 \hangindent=#4
      \Textindent{\the\toks0{\listcount}\the\toks1}}
   \expandafter\expandafter\expandafter
   \edef\c@t#1{begin}{\par
      \countdef\listcount=\the\allocationnumber \listcount=1
      \hangafter=0 \hangindent=#4
      \Textindent{\the\toks0{\listcount}\the\toks1}}
   \expandafter\expandafter\expandafter
   \edef\c@t#1{con}{\par \hangafter=0 \hangindent=#4 \noindent}
   \escapechar=\count255}
\def\c@t#1#2{\csname\string#1#2\endcsname}
\newlist\point=\Number&.&1.0\itemsize;
\newlist\subpoint=(\alphabetic&)&1.75\itemsize;
\newlist\subsubpoint=(\roman&)&2.5\itemsize;
%

%
%
%
\def\keepspacefactor{\let\@sf=\empty \ifhmode
   \edef\@sf{\spacefactor=\the\spacefactor\relax}\relax\fi}
\newcount\footcount \footcount=0
\def\Footnote{\global\advance\footcount by 1 \footnote{\the\footcount}}
\def\footnote#1{\keepspacefactor\refattach{#1}\vfootnote{#1}}

\def\nonfrenchspacing{\sfcode\lq\.=3000 \sfcode\lq\?=3001 \sfcode\lq\!=3001
 \sfcode\lq\:=2000 \sfcode\lq\;=1500 \sfcode\lq\,=1250 }

\nonfrenchspacing
\newcount\referencecount     \referencecount=0
\newif\ifreferenceopen	     \newwrite\referencewrite
\newtoks\rw@toks
\newcount\lastrefsbegincount \lastrefsbegincount=0
\def\refsend{\refmark{\count255=\referencecount
   \advance\count255 by-\lastrefsbegincount
   \ifcase\count255 \number\referencecount
   \or \number\lastrefsbegincount,\number\referencecount
   \else \number\lastrefsbegincount-\number\referencecount \fi}}
\def\refch@ck{\chardef\rw@write=\referencewrite
   \ifreferenceopen \else \referenceopentrue
   \immediate\openout\referencewrite=reference.aux \fi}
%
{\catcode`\^^M=\active 
  \gdef\obeyendofline{\catcode`\^^M\active \let^^M\ }}%
%
{\catcode`\^^M=\active 
  \gdef\ignoreendofline{\catcode`\^^M=5}}
{\obeyendofline\gdef\rw@start#1{\def\t@st{#1}\ifx\t@st\blankend%
\endgroup {\@sf} \relax \else \ifx\t@st\bl@nkend \endgroup {\@sf} \relax%
\else \rw@begin#1
\backtotext
\fi \fi } }
{\obeyendofline\gdef\rw@begin#1
{\def\n@xt{#1}\rw@toks={#1}\relax%
\rw@next}}
\def\blankend{}
{\obeylines\gdef\bl@nkend{
}}
\newif\iffirstrefline  \firstreflinetrue
\def\rwr@teswitch{\ifx\n@xt\blankend \let\n@xt=\rw@begin %
 \else\iffirstrefline \global\firstreflinefalse%
\immediate\write\rw@write{\noexpand\obeyendofline \the\rw@toks}%
\let\n@xt=\rw@begin%
      \else\ifx\n@xt\rw@@d \def\n@xt{\immediate\write\rw@write{%
	\noexpand\ignoreendofline}\endgroup \@sf}%
	     \else \immediate\write\rw@write{\the\rw@toks}%
	     \let\n@xt=\rw@begin\fi\fi \fi}
\def\rw@next{\rwr@teswitch\n@xt}
\def\rw@@d{\backtotext} \let\rw@end=\relax
\let\backtotext=\relax

\newdimen\refindent	\refindent=20pt
\newmuskip\refskip
\newmuskip\regularrefskip \regularrefskip=2mu
\newmuskip\specialrefskip \specialrefskip=-2mu
\def\refattach#1{\@sf \ifhmode\ifnum\spacefactor=1250 \refskip=\specialrefskip
 \else\ifnum\spacefactor=3000 \refskip=\specialrefskip
 \else\ifnum\spacefactor=1001 \refskip=\specialrefskip
 \else \refskip=\regularrefskip \fi\fi\fi
 \else \refskip=\regularrefskip \fi
 \ref@ttach{\strut^{\mkern\refskip #1}}}
\def\ref@ttach#1{\ifmmode #1\else\ifhmode\unskip${}#1$\relax\fi\fi{\@sf}}
\def\PLrefmark#1{ [#1]{\@sf}}
\def\NPrefmark#1{\refattach{\scriptstyle [ #1 ] }}
\let\PRrefmark=\refattach
\def\refmark{\keepspacefactor\refm@rk}
\def\refm@rk#1{\relax\therefm@rk{#1}}
\def\originalrefs{\let\therefm@rk=\NPrefmark}
\def\PRrefs{\let\therefm@rk=\PRrefmark \let\therefitem=\PRrefitem}
\def\PLrefs{\let\therefm@rk=\PLrefmark \let\therefitem=\PLrefitem}
\def\PRrefitem#1{\refitem{#1.}}
\def\PLrefitem#1{\refitem{[#1]}}
\let\therefitem=\PRrefitem
\def\refitem#1{\par \hangafter=0 \hangindent=\refindent \Textindent{#1}}
\def\REFNUM#1{\eatspace\keepspacefactor\refch@ck \firstreflinetrue%
 \global\advance\referencecount by 1 \xdef#1{\the\referencecount}}
\def\eatspace{\ifhmode\unskip\fi}
\def\refnum#1{\keepspacefactor\refch@ck \firstreflinetrue%
 \global\advance\referencecount by 1 \xdef#1{\the\referencecount}\refend}
\def\REF#1{\REFNUM#1%
 \immediate\write\referencewrite{%
 \noexpand\therefitem{#1}}%
\begingroup\obeyendofline\rw@start}
\def\ref{\refnum\?%
 \immediate\write\referencewrite{\noexpand\therefitem{\?}}%
\begingroup\obeyendofline\rw@start}
\def\Ref#1{\refnum#1%
 \immediate\write\referencewrite{\noexpand\therefitem{#1}}%
\begingroup\obeyendofline\rw@start}
\def\REFS#1{\REFNUM#1\global\lastrefsbegincount=\referencecount
\immediate\write\referencewrite{\noexpand\therefitem{#1}}%
\begingroup\obeyendofline\rw@start}
\def\refend{\refm@rk{\number\referencecount}}
%

%

\newcount\figurecount	  \figurecount=0
\newif\iffigureopen	  \newwrite\figurewrite
\newdimen\digitwidth \setbox0=\hbox{\rm0} \digitwidth=\wd0
\def\zerophant{\kern\digitwidth}
\def\FIGNUM#1{\keepspacefactor\figch@ck \firstreflinetrue%
\global\advance\figurecount by 1 \xdef#1{\the\figurecount}}
\def\figch@ck{\chardef\rw@write=\figurewrite \iffigureopen\else
   \immediate\openout\figurewrite=figures.aux
   \figureopentrue\fi}
\def\figitem#1{\par\indent \hangindent2\parindent \textindent{Fig. #1\ }}
\def\FIGLABEL#1{\ifnum\number#1<10 \def\figlabel{#1.\zerophant}\else%
\def\figlabel{#1.}\fi}
\def\FIG#1{\FIGNUM#1\FIGLABEL#1%
\immediate\write\figurewrite{\noexpand\figitem{\figlabel}}%
\begingroup\obeyendofline\rw@start}
\def\Figname#1{\FIGNUM#1Fig.~#1\FIGLABEL#1%
\immediate\write\figurewrite{\noexpand\figitem{\figlabel}}%
\begingroup\obeyendofline\rw@start}
\def\fig{\FIGNUM\? fig.~\? \FIGLABEL\?
\immediate\write\figurewrite{\noexpand\figitem{\figlabel}}%
\begingroup\obeyendofline\rw@start}
\def\figure{\FIGNUM\? figure~\? \FIGLABEL\?
\immediate\write\figurewrite{\noexpand\figitem{\figlabel}}%
\begingroup\obeyendofline\rw@start}
\def\Fig{\FIGNUM\? Fig.~\? \FIGLABEL\?
\immediate\write\figurewrite{\noexpand\figitem{\figlabel}}%
\begingroup\obeyendofline\rw@start}
\def\Figure{\FIGNUM\? Figure~\? \FIGLABEL\?
\immediate\write\figurewrite{\noexpand\figitem{\figlabel}}%
\begingroup\obeyendofline\rw@start}
\def\figout{\par \penalty-400 \vskip\chapterskip
  \spacecheck\referenceminspace \immediate\closeout\figurewrite
  \figureopenfalse
  \line{\fourteenrm
   \hfil FIGURE CAPTION\ifnum\figurecount=1 \else S \fi\hfil}
  \vskip\headskip
  \input figures.aux
  }
\newcount\tablecount	 \tablecount=0
\newif\iftableopen	 \newwrite\tablewrite
\def\tabch@ck{\chardef\rw@write=\tablewrite \iftableopen\else
   \immediate\openout\tablewrite=tables.aux
   \tableopentrue\fi}
\def\TABNUM#1{\keepspacefactor\tabch@ck \firstreflinetrue%
\global\advance\tablecount by 1 \xdef#1{\the\tablecount}}
\def\tableitem#1{\par\indent \hangindent2\parindent \textindent{Table #1\ }}
\def\TABLE#1{\TABNUM#1\FIGLABEL#1%
\immediate\write\tablewrite{\noexpand\tableitem{\figlabel}}%
\begingroup\obeyendofline\rw@start}
\def\Table{\TABNUM\? Table~\?\FIGLABEL\?%
\immediate\write\tablewrite{\noexpand\tableitem{\figlabel}}%
\begingroup\obeyendofline\rw@start}
\def\tabout{\par \penalty-400 \vskip\chapterskip
  \spacecheck\referenceminspace \immediate\closeout\tablewrite \tableopenfalse
  \line{\fourteenrm\hfil TABLE CAPTION\ifnum\tablecount=1 \else S\fi\hfil}
  \vskip\headskip
  \input tables.aux
  }
\PRrefs
\def\etal{{\it et al.}}
%
\def\masterreset{\global\pagenumber=1 \global\chapternumber=0
   \global\equanumber=0 \global\sectionnumber=0
   \global\referencecount=0 \global\figurecount=0 \global\tablecount=0 }
\def\FRONTPAGE{\ifvoid255\else\vfill\penalty-2000\fi
      \masterreset\global\frontpagetrue
      \global\lastf@@t=0 \global\footsymbolcount=0}

\def\papersize{\hsize=35pc\vsize=50pc\hoffset=1pc\voffset=6pc
  \skip\footins=\bigskipamount}
\def\paperstyle{\normalspace\papersize}
\paperstyle
\newskip\frontpageskip
\newtoks\Pubnum \newtoks\pubnum
\newtoks\s@condpubnum \newtoks\th@rdpubnum
\newif\ifs@cond \s@condfalse
\newif\ifth@rd \th@rdfalse
\newif\ifp@bblock  \p@bblocktrue
\newcount\Year
\def\Yearset{\Year=\year \advance\Year by -1900
 \ifnum\month<4 \advance\Year by -1 \fi}
\def\PH@SR@V{\doubl@true \baselineskip=24.1pt plus 0.2pt minus 0.1pt
	     \parskip= 3pt plus 2pt minus 1pt }
\def\PHYSREV{\paperstyle\PhysRevtrue\PH@SR@V}
\def\titlepage{\Yearset\FRONTPAGE\paperstyle\ifPhysRev\PH@SR@V\fi
   \ifp@bblock\p@bblock\fi}
\def\nopubblock{\p@bblockfalse}
\def\endpage{\vfil\break}
\frontpageskip=1\medskipamount plus .5fil
\Pubnum={TU--\the\pubnum }
\pubnum={ }
\def\secondpubnum#1{\s@condtrue\s@condpubnum={#1}}
\def\thirdpubnum#1{\th@rdtrue\th@rdpubnum={#1}}
\def\p@bblock{\begingroup \tabskip=\hsize minus \hsize
   \baselineskip=1.5\ht\strutbox \topspace-2\baselineskip
   \halign to\hsize{\strut ##\hfil\tabskip=0pt\crcr
   \the\Pubnum\cr
   \ifs@cond \the\s@condpubnum\cr\fi
   \ifth@rd \the\th@rdpubnum\cr\fi
   \the\Month \cr}\endgroup}
\def\title#1{\hrule height0pt depth0pt
   \vskip\frontpageskip \titlestyle{#1} \vskip\headskip }
\def\author#1{\vskip\frontpageskip\titlestyle{\twelvecp #1}\nobreak}

\def\address#1{\par\kern 5pt\titlestyle{\twelvepoint\it #1}}
\def\andaddress{\par\kern 5pt \centerline{\sl and} \address}
\def\abstract{\vskip\frontpageskip\centerline{\fourteenrm ABSTRACT}
 \vskip\headskip }

%
%
%
     \def\etc{\hbox{\it etc.}}

\def\\{\relax\ifmmode\backslash\else$\backslash$\fi}
\def\globaleqnumbers{\relax\if\equanumber<0\else\global\equanumber=-1\fi}
\def\nextline{\unskip\nobreak\hskip\parfillskip\break}

\def\topspace{\hrule height 0pt depth 0pt \vskip}

\let\int=\intop 
\def\prop{\mathrel{{\mathchoice{\pr@p\scriptstyle}{\pr@p\scriptstyle}%
 {\pr@p\scriptscriptstyle}{\pr@p\scriptscriptstyle} }}}
\def\pr@p#1{\setbox0=\hbox{$\cal #1 \char'103$}
   \hbox{$\cal #1 \char'117$\kern-.4\wd0\box0}}
\def\lsim{\mathrel{\mathpalette\@versim<}}
\def\gsim{\mathrel{\mathpalette\@versim>}}
\def\@versim#1#2{\lower0.2ex\vbox{\baselineskip\z@skip\lineskip\z@skip
  \lineskiplimit\z@\ialign{$\m@th#1\hfil##\hfil$\crcr#2\crcr\sim\crcr}}}
%
%
%
\let\sec@nt=\sec
\def\sec{\relax\ifmmode\let\n@xt=\sec@nt\else\let\n@xt\section\fi\n@xt}
\def\obsolete#1{\message{Macro \string #1 is obsolete.}}
\def\firstsec#1{\obsolete\firstsec \section{#1}}
\def\firstsubsec#1{\obsolete\firstsubsec \subsection{#1}}
\def\thispage#1{\obsolete\thispage \global\pagenumber=#1\frontpagefalse}
\def\thischapter#1{\obsolete\thischapter \global\chapternumber=#1}
\def\nextequation#1{\obsolete\nextequation \global\equanumber=#1
   \ifnum\the\equanumber>0 \global\advance\equanumber by 1 \fi}
\def\BOXITEM{\afterassigment\B@XITEM\setbox0=}
\def\B@XITEM{\par\hangindent\wd0 \noindent\box0 }
%

%
\catcode`\@=12
%


\def\jnfont{\rm}
\def\APPB#1,{{\jnfont Acta Phys.\ Polon.}\ {\bf B#1},}
\def\AP#1,{{\jnfont Ann.\ Phys.\ (N.Y.)} {\bf #1},}
\def\ApJ#1,{{\jnfont Astrophys.\ J.}\ {\bf #1},}
\def\EpL#1,{{\jnfont Europhys.\ Lett.}\ {\bf #1},}
\def\IJMPA#1,{{\jnfont Int.\ J.\ Mod.\ Phys.\ A}~{\bf #1},}
\def\JETP#1,{{\jnfont Sov.\ Phys.\ JETP}\ {\bf #1},}
\def\JETPL#1,{{\jnfont JETP Lett.}\ {\bf #1},}
\def\JMP#1,{{\jnfont J. Math.\ Phys.}\ {\bf #1},}
\def\LNC#1,{{\jnfont Lett.\ Nuovo Cimento} {\bf #1},}
\def\MPLA#1,{{\jnfont Mod.\ Phys.\ Lett.\ A}~{\bf #1},}
\def\NC#1,{{\jnfont Nuovo Cimento} {\bf #1},}
\def\NP#1,{{\jnfont Nucl.\ Phys.}\ {\bf #1},}
\def\NPA#1,{{\jnfont Nucl.\ Phys.}\ {\bf A#1},}
\def\NPB#1,{{\jnfont Nucl.\ Phys.}\ {\bf B#1},}
\def\Physica#1,{{\jnfont Physica}\ {\bf #1},}
\def\PL#1,{{\jnfont Phys.\ Lett.}\ {\bf #1},}
\def\PLB#1,{{\jnfont Phys.\ Lett.\ B}~{\bf #1},}
\def\PRep#1,{{\jnfont Phys.\ Rep.}\ {\bf #1},}
\def\PR#1,{{\jnfont Phys.\ Rev.}\ {\bf #1},}
\def\PRD#1,{{\jnfont Phys.\ Rev.\ D}~{\bf #1},}
\def\PRL#1,{{\jnfont Phys.\ Rev.\ Lett.}\ {\bf #1},}
\def\PTP#1,{{\jnfont Prog.\ Theor.\ Phys.}\ {\bf #1},}
\def\PZETF#1,{{\jnfont Pis'ma Zh.\ Eksp.\ Teor.\ Fiz.}\ {\bf #1},}
\def\RMP#1,{{\jnfont Rev.\ Mod.\ Phys.}\ {\bf #1},}
\def\SJNP#1,{{\jnfont Sov.\ J. Nucl.\ Phys.}\ {\bf #1},}
\def\YaF#1,{{\jnfont Yad.\ Fiz.}\ {\bf #1},}
\def\ZETF#1,{{\jnfont Zh.\ Eksp.\ Teor.\ Fiz.}\ {\bf #1},}
\def\ZPC#1,{{\jnfont Z. Phys.\ C} {\bf #1},}
\font\fourteenbi=cmmib10 scaled\magstep2   \skewchar\fourteenbi='177
\font\twelvebi=cmmib10 scaled\magstep1     \skewchar\twelvebi='177
\font\elevenbi=cmmib10 scaled\magstephalf  \skewchar\elevenbi='177
\font\tenbi=cmmib10                        \skewchar\tenbi='177
\font\fourteenbsy=cmbsy10 scaled\magstep2  \skewchar\fourteenbsy='60
\font\twelvebsy=cmbsy10 scaled\magstep1    \skewchar\twelvebsy='60
\font\elevenbsy=cmbsy10 scaled\magstephalf \skewchar\elevenbsy='60
\font\tenbsy=cmbsy10                       \skewchar\tenbsy='60
\font\fourteenbsl=cmbxsl10 scaled\magstep2
\font\twelvebsl=cmbxsl10 scaled\magstep1
\font\elevenbsl=cmbxsl10 scaled\magstephalf
\font\tenbsl=cmbxsl10
\font\fourteenbit=cmbxti10 scaled\magstep2
\font\twelvebit=cmbxti10 scaled\magstep1
\font\elevenbit=cmbxti10 scaled\magstephalf
\font\tenbit=cmbxti10
\catcode\lq\@=11
\def\fourteenbold{\relax
    \textfont0=\fourteenbf	    \scriptfont0=\tenbf
    \scriptscriptfont0=\sevenbf
     \def\rm{\fam0 \fourteenbf \f@ntkey=0 }\relax
    \textfont1=\fourteenbi	    \scriptfont1=\tenbi
    \scriptscriptfont1=\seveni
     \def\oldstyle{\fam1 \fourteenbi\f@ntkey=1 }\relax
    \textfont2=\fourteenbsy	    \scriptfont2=\tenbsy
    \scriptscriptfont2=\sevensy
    \textfont3=\fourteenex     \scriptfont3=\fourteenex
    \scriptscriptfont3=\fourteenex
    \def\it{\fam\itfam \fourteenbit\f@ntkey=4 }\textfont\itfam=\fourteenbit
    \def\sl{\fam\slfam \fourteenbsl\f@ntkey=5 }\textfont\slfam=\fourteenbsl
    \scriptfont\slfam=\tensl
    \def\bf{\fam\bffam \fourteenrm\f@ntkey=6 }\textfont\bffam=\fourteenrm
    \scriptfont\bffam=\tenrm	 \scriptscriptfont\bffam=\sevenrm
    \def\tt{\fam\ttfam \twelvett \f@ntkey=7 }\textfont\ttfam=\twelvett
    \h@big=11.9\p@ \h@Big=16.1\p@ \h@bigg=20.3\p@ \h@Bigg=24.5\p@
    \def\caps{\fam\cpfam \twelvecp \f@ntkey=8 }\textfont\cpfam=\twelvecp
    \setbox\strutbox=\hbox{\vrule height 12pt depth 5pt width\z@}\relax
    \samef@nt}
\def\twelvebold{\relax
    \textfont0=\twelvebf	  \scriptfont0=\ninebf
    \scriptscriptfont0=\sevenbf
     \def\rm{\fam0 \twelvebf \f@ntkey=0 }\relax
    \textfont1=\twelvebi	  \scriptfont1=\ninei
    \scriptscriptfont1=\seveni
     \def\oldstyle{\fam1 \twelvebi\f@ntkey=1 }\relax
    \textfont2=\twelvebsy	  \scriptfont2=\ninesy
    \scriptscriptfont2=\sevensy
    \textfont3=\twelveex	  \scriptfont3=\twelveex
    \scriptscriptfont3=\twelveex
    \def\it{\fam\itfam \twelvebit \f@ntkey=4 }\textfont\itfam=\twelvebit
    \def\sl{\fam\slfam \twelvebsl \f@ntkey=5 }\textfont\slfam=\twelvebsl
    \scriptfont\slfam=\ninesl
    \def\bf{\fam\bffam \twelverm \f@ntkey=6 }\textfont\bffam=\twelverm
    \scriptfont\bffam=\ninerm	  \scriptscriptfont\bffam=\sevenrm
    \def\tt{\fam\ttfam \twelvett \f@ntkey=7 }\textfont\ttfam=\twelvett
    \h@big=10.2\p@ \h@Big=13.8\p@ \h@bigg=17.4\p@ \h@Bigg=21.0\p@
    \def\caps{\fam\cpfam \twelvecp \f@ntkey=8 }\textfont\cpfam=\twelvecp
    \setbox\strutbox=\hbox{\vrule height 10pt depth 4pt width\z@}\relax
    \samef@nt}
\def\elevenbold{\relax
    \textfont0=\elevenbf	  \scriptfont0=\ninebf
    \scriptscriptfont0=\sixbf
     \def\rm{\fam0 \elevenbf \f@ntkey=0 }\relax
    \textfont1=\elevenbi	  \scriptfont1=\ninei
    \scriptscriptfont1=\sixi
     \def\oldstyle{\fam1 \elevenbi\f@ntkey=1 }\relax
    \textfont2=\elevenbsy	  \scriptfont2=\ninesy
    \scriptscriptfont2=\sixsy
    \textfont3=\elevenex	  \scriptfont3=\elevenex
    \scriptscriptfont3=\elevenex
    \def\it{\fam\itfam \elevenbit \f@ntkey=4 }\textfont\itfam=\elevenbit
    \def\sl{\fam\slfam \elevenbsl \f@ntkey=5 }\textfont\slfam=\elevenbsl
    \scriptfont\slfam=\ninesl
    \def\bf{\fam\bffam \elevenrm \f@ntkey=6 }\textfont\bffam=\elevenrm
    \scriptfont\bffam=\ninerm	  \scriptscriptfont\bffam=\sixrm
    \def\tt{\fam\ttfam \eleventt \f@ntkey=7 }\textfont\ttfam=\eleventt
    \h@big=9.311\p@ \h@Big=12.6\p@ \h@bigg=15.88\p@ \h@Bigg=19.17\p@
    \def\caps{\fam\cpfam \elevencp \f@ntkey=8 }\textfont\cpfam=\elevencp
    \setbox\strutbox=\hbox{\vrule height 9pt depth 4pt width\z@}\relax
    \samef@nt}
\def\tenbold{\relax
    \textfont0=\tenbf	       \scriptfont0=\sevenrm
    \scriptscriptfont0=\fiverm
    \def\rm{\fam0 \tenrm \f@ntkey=0 }\relax
    \textfont1=\tenbi	       \scriptfont1=\seveni
    \scriptscriptfont1=\fivei
    \def\oldstyle{\fam1 \tenbi \f@ntkey=1 }\relax
    \textfont2=\tenbsy	       \scriptfont2=\sevensy
    \scriptscriptfont2=\fivesy
    \textfont3=\tenex	       \scriptfont3=\tenex
    \scriptscriptfont3=\tenex
    \def\it{\fam\itfam \tenbit \f@ntkey=4 }\textfont\itfam=\tenbit
    \def\sl{\fam\slfam \tenbsl \f@ntkey=5 }\textfont\slfam=\tenbsl
    \def\bf{\fam\bffam \tenrm \f@ntkey=6 }\textfont\bffam=\tenrm
    \scriptfont\bffam=\sevenrm  \scriptscriptfont\bffam=\fiverm
    \def\tt{\fam\ttfam \tentt \f@ntkey=7 }\textfont\ttfam=\tentt
    \def\caps{\fam\cpfam \tencp \f@ntkey=8 }\textfont\cpfam=\tencp
    \h@big=8.5\p@ \h@Big=11.5\p@ \h@bigg=14.5\p@ \h@Bigg=17.5\p@
    \setbox\strutbox=\hbox{\vrule height 8.5pt depth 3.5pt width\z@}\relax
    \samef@nt}
\def\bold{\iftwelv@\twelvebold\else\ifelev@n\elevenbold\else\tenbold\fi\fi}
\catcode\lq\@=12
\let\to=\rightarrow
\foottokens={\vskip 1\jot\Tenpoint }

\def\r{\noalign{\vskip 3\jot }}

\def\tfrac#1#2{{\textstyle{#1\over #2}}}

\def\({\hbox{\sevenrm(}\mskip-1mu}
\def\){\hbox{\sevenrm)}}

\mathchardef\REAL="023C
\mathchardef\IMAG="023D

\def\hidehrule#1#2{\kern-#1 \hrule height#1 depth#2 \kern-#2 }
\def\hidevrule#1#2{\kern-#1{\dimen0=#1 \advance\dimen0 by #2%
\vrule width\dimen0}\kern-#2 }
\def\makeblankbox#1#2{\hbox{\lower\dp0\vbox{\hidehrule{#1}{#2}%
\kern-#1 \hbox to \wd0{\hidevrule{#1}{#2}\raise\ht0\vbox to #1{}%
\lower\dp0\vtop to #1{}\hfil\hidevrule{#2}{#1}}\kern -#1\hidehrule{#2}{#1}}}}
\def\dAlembert{\setbox0=\hbox{$\Sigma$}\kern .1em \lower .1ex
\makeblankbox{.25pt}{.25pt}\kern .1em }

{\catcode`\|=0 \catcode`\\=12 
  |obeylines|gdef|doverbatim^^M#1\endverbatim{#1|endgroup}}

 {\obeyspaces\global\let =\ } 

\def\goodmood{\hbox{$\rlap{\kern.35ex$\scriptscriptstyle\smile
$}\rlap{\kern.55ex\lower.45ex\hbox{$\mathchar"707F$}}\bigcirc$}}
\def\badmood{\hbox{$\rlap{\kern.35ex$\scriptscriptstyle\frown
$}\rlap{\kern.55ex\lower.45ex\hbox{$\mathchar"707F$}}\bigcirc$}}
\def\verybadmood{\hbox{$\rlap{\kern.5ex\lower1.25ex\hbox{$\mathchar"0365
$}}\rlap{\kern.55ex\lower.45ex\hbox{$\mathchar"707F$}}\bigcirc$}}
\def\leftmood{\hbox{$\rlap{\kern.5ex\lower1.25ex\hbox{$\mathchar"0365
$}}\rlap{\kern.35ex\lower.45ex\hbox{$\mathchar"707F$}}\bigcirc$}}
\def\rightmood{\hbox{$\rlap{\kern.5ex\lower1.25ex\hbox{$\mathchar"0365
$}}\rlap{\kern.75ex\lower.45ex\hbox{$\mathchar"707F$}}\bigcirc$}}
%
\def\GENITEM#1;#2{\par \hangafter=0 \hangindent=#1
    \Textindent{#2}\ignorespaces}
%
%
\newitem\item=1\itemsize;
\newitem\sitem=1.75\itemsize;	  
\newitem\ssitem=2.5\itemsize;	  
\newitem\appitem=2.9\itemsize;
%
\font\seventeenbi=cmmib10 scaled\magstep3
\font\twelvebi=cmmib10 scaled\magstep1
\font\twelvebsy=cmbsy10 scaled\magstep1
\def\title#1{\hrule height0pt depth0pt
   \vskip\frontpageskip \titlestyle{\baselineskip=35pt \begingroup
   \textfont1=\seventeenbi \scriptfont1=\twelvebi
   \scriptfont0=\twelvebf \scriptfont2=\twelvebsy
   \seventeenbf #1 \endgroup} \vskip\headskip }
\def\author#1{\vskip\frontpageskip\titlestyle{\fourteenrm #1}\nobreak}
\font\seventeenbi=cmmib10 scaled\magstep3
\advance\vsize by 36pt
\advance\voffset by -36pt
\pubnum{475}
\Month{January 1995}
\PLrefs
%
\finaltrue

\catcode`\@=13 \def@{{\vphantom.}}
\def\frac#1#2{{#1\over #2}}
\def\sq{\tilde q}
\def\gluino{\tilde g}
\def\photino{\mathchoice{\widetilde\gamma}{\widetilde\gamma}%
 {\tilde\gamma}{\tilde\gamma}}
\def\neut{\mathchoice{\widetilde\chi}{\widetilde\chi}%
 {\tilde\chi}{\tilde\chi}^0}
\def\PLeft{{\textstyle{1-\gamma_5\over2}}}
\def\PR{{\textstyle{1+\gamma_5\over2}}}
\def\nL{n^{\scriptscriptstyle(L)}}
\def\nR{n^{\scriptscriptstyle(R)}}
\def\Li#1(#2){\mathop{\rm Li}\nolimits_{#1}\left({#2}\right)}
%

\def\chapter#1{\par \penalty-300 \vskip\chapterskip
   \spacecheck\chapterminspace
   \chapterreset \par\noindent{\twelvebold\chapterlabel \ #1}\par
   \nobreak\vskip\headskip \penalty 30000
   \wlog{\string\chapter\ \chapterlabel} }
\def\section#1{\par \ifnum\the\lastpenalty=30000\else
   \penalty-200\vskip\sectionskip \spacecheck\sectionminspace\fi
   \wlog{\string\section\ \chapterlabel \the\sectionnumber}
   \global\advance\sectionnumber by 1  \noindent
   {{\elevenbold #1}}\par
   \nobreak\vskip\headskip \vskip-\parskip \penalty 30000 }
\chapterminspace=120pt
\sectionminspace=60pt

\def\figout{\par \penalty-400 \vskip\chapterskip
   \spacecheck\referenceminspace \immediate\closeout\figurewrite
   \figureopenfalse
   \leftline{\twelvebold Figure Captions}\par
   \nobreak\vskip\headskip \penalty 30000
   \input figures.aux
   }
\newcount\footnotecount \footnotecount=0
\let\ofootnote=\footnote
\def\footnote{\global\advance\footnotecount by 1
 \ofootnote{\the\footnotecount}}
\titlepage
\title{\begingroup\textfont1=\seventeenbi \scriptfont1=\twelvebi
\scriptfont0=\twelvebf \scriptfont2=\twelvebsy
\seventeenbf Soft-breaking correction to hard supersymmetric relations
\break
\fourteenbf ---QCD correction to squark decay---\endgroup}
\author{\fourteenrm  Ken-ichi Hikasa and Yumi Nakamura}
\address{Physics Department, Tohoku University\break
Aoba-ku, Sendai 980-77, Japan}

\abstract

Supersymmetric relations between dimensionless couplings receive finite
correction at one-loop when supersymmetry is broken softly.  We calculate
the ${\cal O}(\alpha_s)$ correction to the squark decay width to a quark
and an electroweak gaugino, which is found to be nonvanishing.
Logarithmic correction appears when the gluino is heavy.

\endpage
\pagenumber=1
\sequentialequations

\chapter{Introduction}

Supersymmetric field theories comprise a very special subset of general
field theories.  First of all, a field theory can be supersymmetric
only when the number of bosonic and fermionic fields are the same.
Supersymmetry thus predicts the existence of superpartners.
Moreover, interactions are tightly interrelated
if a theory should possess supersymmetry.
For example, a selectron couples to a photino and an electron
with the coupling strength given by the electromagnetic gauge coupling.
If the nature really possesses supersymmetry, discovery of superparticles
is thus only the first step to prove it.  Verification of supersymmetric
relations between various couplings is necessary to establish the theory.

In the minimal supersymmetric standard model (MSSM), supersymmetry is
softly broken such that no quadratic divergences appear in mass terms
and tadpoles.  The breaking consists\ref{L.~Girardello and M.~T. Grisaru,
\NPB194, 65 (1982).} of the scalar and gaugino mass terms and a certain
type of three-scalar couplings (the so-called $A$ term).
Superpartners receive a mass of the order of the weak scale from the
breaking.  Meanwhile, dimension-four interaction terms are not modified, so
the selectron-electron-photino coupling retains the value $e$.

When loop effects are included, however, the soft breaking affects
the dimen\-sion-four couplings.
The general theory of renormalization%
\ref{K.~Symanzik, in {\it Fundamental Interactions at High Energies},
edited by\hfil\break  A.~Perlmutter \etal\ (Gordon and Breach, 1970).}
states that soft symmetry breaking does not generate a new divergence
in dimension-four vertices.  No new counterterm is called for.%
\footnote{Soft breaking of supersymmetry is soft in the renormalization
theory sense, though the opposite is not true.}
Nevertheless, the equality of the couplings prescribed by supersymmetry
receives {\it finite\/} modification.

In this paper, we examine this effect in a simple example:
${\cal O}(\alpha_s)$ correction to the squark-quark-photino coupling.
Physically, this vertex can be measured as the decay width of the squark.
At the tree level, this width is expressed in terms of the electromagnetic
coupling.  We will find that there is indeed a finite ${\cal O}(\alpha_s)$
correction to the width.  In contrast, the coupling receives no modification
if supersymmetry is exact.

Numerous works have been done on calculation of radiative corrections in
MSSM, but they are limited to the effects of supersymmetric particles to
processes governed by the gauge couplings.
To our knowledge, no calculation exists on corrections to hard
supersymmetric relations.

Although we refer to photinos most of the time, our result for the
correction factor applies to squark decay to a quark and any
electroweak gaugino (neutralino or chargino).

\chapter{Lowest-order coupling}

The squark-quark-neutralino coupling in MSSM is given by%
\footnote{Here we neglect the Yukawa-type interaction
of the higgsino component of the neutralino, which is proportional to the
quark mass.}
$${\cal L}= -\sqrt2 e \Bigl( \nL_i  \bar q \PR \neut_i \sq_L
            - \nR_i \bar q \PLeft \neut_i \sq_R  \Bigr) + \hbox{h.c.}
\;,\Eqn\eqInt$$
where
$$\eqalignno{
\nL_i &= \bigl[ T_{3L} N_{iz} / \!\sin\theta_W
 + (Q-T_{3L}) N_{ib} / \!\cos\theta_W \bigr] \;,&\Eqa\cr
\nR_i &=  Q N_{ib}^* / \!\cos\theta_W  \;,&\Eqb\cr}$$
and $N_{iz}$, $N_{ib}$ are the neutralino mixing matrix elements:
$$ \neut_{iL} = N_{ib} \widetilde B_L + N_{iz} \widetilde Z_L
+ \hbox{Higgsino components}. \Eq$$
For a photino
($N_{\photino b}=\cos\theta_W$, $N_{\photino z} = \sin\theta_W$),
one has $\nL_{\photino} = \nR_{\photino} = Q$.
Supersymmetry thus constrains the squark-quark-photino ``Yukawa'' coupling
to be equal (up to a ``Clebsch-Gordan'' constant) to the electromagnetic
gauge coupling $e$ at the tree order, even with soft breaking.

The squark decay width to a quark and a neutralino is found to be
$$\Gamma_0(\sq_L\to q\neut_i)= \tfrac12\alpha |\nL_i|^2 m_{\sq} (1-r)^2 \;,
\Eq$$
where $r=m_{\neut}^2/m_{\sq}^2$.  For a photino
$$\Gamma_0(\sq\to q\photino)= \tfrac12\alpha Q^2 m_{\sq} (1-r)^2 \;.\Eq$$
The width is determined by the electromagnetic gauge coupling.

\chapter{Comment on the Computational Method}

Although supergraph method is powerful in computations with exact
supersymmetry,  its use seems to be limited in a softly broken theory.%
\REF\YY{See e.g., Y.~Yamada, \PRD50, 3537 (1994).}%
\footnote{The spurion technique to include soft breaking is useful in
the calculation of divergent quantities like beta functions\refmark{\YY}, or
in situations in which the soft breaking can be treated as perturbation.}
We believe that ordinary Feynman graph technique is more convenient.
However, there are several complications in practice.

First, supersymmetry should not be violated by regularization.  No
method is known which fully respects supersymmetry.  We use the dimensional
reduction method\ref{W.~Siegel, \PL84B, 193 (1979);\nextline
D.~M. Capper, D.R.T. Jones, and P. van Nieuwenhuisen, \NPB167, 479
(1980);\nextline
L.~V. Avdeev and A.~A. Vladimirov, \NPB219, 262 (1983);\nextline
P.~Howe, A.~Parkes, and P.~West, \PL147B, 409 (1984).}
which is compatible with supersymmetry at least at one-loop order.

Second, manifest supersymmetry is lost when we fix the gauge.  We work in
Wess-Zumino gauge to remove some unphysical fields in the gauge
superfield, and further use Feynman gauge to define the gauge field
propagator.   This has the consequence%
\ref{J.~Wess and B.~Zumino, \NPB78, 1 (1974).}
that the wave function renormalization constant (even the divergent part)
for the scalar and the fermion differ from each other.  One of its
implications is that the usual supersymmetric transformation
rule does not hold for the renormalized matter field.
A care is thus needed in determining how to renormalize the ultraviolet
divergence in the squark-quark-photino vertex.  Without supersymmetry,
this interaction would be an independent coupling on which
one could set up any renormalization condition at will.
In fact, supersymmetry prescribes the counterterm for the vertex
which removes the divergence.

\chapter{${\cal O}(\alpha_s)$ correction under exact supersymmetry}

Before discussing the ${\cal O}(\alpha_s)$ correction to the squark
decay width, we explicitly demonstrate that the equality of $\bar qq\gamma$
and $\bar q \sq\photino$ coupling is
not modified at ${\cal O}(\alpha_s)$ if supersymmetry is not broken.
We assume the quark and squark have the same mass $m_q\neq0$,
and the gluon, gluino, photon, and photino are massless%
\footnote{Here the quark and squark can be thought to have gauge-invariant
masses, because only strong and electromagnetic couplings enter at the order
we work.  The corrections discussed below are thus identical for $\sq_L$
and $\sq_R$.}

The one-loop graphs for the $\bar q\sq\photino$ coupling is shown
in Fig.~1.  We evaluate these diagrams at the ``on-shell'' limit:
the quark (squark) are on their mass shell, and the four-momentum squared
$q^2$ of the photino is taken to be $q^2\to 0$.  For the $\bar qq\gamma$
vertex n the corresponding limit, there is no ${\cal O}(\alpha_s)$
correction in the on-shell renormalization scheme.  We use dimensional
reduction with $D=4-2\epsilon$ for ultraviolet cutoff and regularize
infrared divergences by an infinitesimal gluon mass $\lambda$.
The gluon exchange diagram gives
$${C_F\alpha_s\over4\pi}\biggl[{1\over\epsilon} - \log{m^2\over\mu^2}
- 2\log{m^2\over\lambda^2}+4 \biggr]\times(\rm lowest) \;,\Eqn\eqgluon$$
and the gluino exchange contributes
$${C_F\alpha_s\over4\pi}\cdot 2 \times(\rm lowest) \;.\Eqn\eqgluino$$
Here $C_F=4/3$ is a color factor
and $\mu$ is the arbitrary renormalization scale.
\footnote{In the usual convention, our $1/\epsilon$
should be read as $1/\epsilon-\gamma_E^@ + \ln4\pi$.}
The sum of these contribution is both ultraviolet and infrared divergent.
The necessary counterterm to render it finite may be found as follows.
At ${\cal O}(\alpha_s)$, neither the QED coupling $e$ nor the photino field
receives corrections.  The counterterm is then determined by the wave
function renormalization for the quark and squark fields to be
$$\bigl( Z_q^{1/2} Z_{\sq}^{1/2} -1 \bigr) \times
\hbox{(lowest order vertex)} \;.\Eqn\eqcounterterm$$
Here $Z_q$ ($Z_{\sq}$) is the wave function renormalization constant for
the quark (squark).
We evaluate the renormalization constants in the on-shell renormalization
scheme from the quark and squark two-point functions.  The diagrams
needed are shown in Fig.~2.
We find
$$\eqalignno{Z_q-1 &= {C_F\alpha_s\over4\pi}\biggl[-{2\over\epsilon}
+ 2\log{m^2\over\mu^2} + 2\log{m^2\over\lambda^2} - 8 \biggr] \;,&\Eqa\cr\r
Z_{\tilde q}-1 &= {C_F\alpha_s\over4\pi}\biggl[
 2\log{m^2\over\lambda^2} - 4 \biggr] \;,&\Eqb\cr}$$
The two counterterms are not equal because our calculation is
in the Wess-Zumino-Feynman gauge which is not manifestly symmetric.%
\footnote{The mass counterterms are found to satisfy the manifestly
supersymmetric relation $\delta m_{\sq}^2 = 2m_q\delta m_q$.}
The counterterm contribution \eqcounterterm\ exactly cancels the
one-loop graphs \eqgluon\ and \eqgluino.
Therefore, the supersymmetric relation between the $\bar qq\gamma$
and $\bar q\sq\photino$ couplings receives no correction at
${\cal O}(\alpha_s)$ when supersymmetry is exact.%
\REF\Tobe{K.~Tobe, master thesis (in Japanese), Tohoku University (1994).}%
\footnote{This result can be verified using supersymmetric Ward identity%
\refmark{\Tobe}.}

\chapter{${\cal O}(\alpha_s)$ correction to the squark decay width}

Now we turn on the soft supersymmetry breaking which shifts upward the mass
of the squark, gluino, and photino.  We assume $m_{\photino}<m_{\sq}$
so that the on-shell process $\sq\to q\photino$ is kinematically allowed.
Although we consider the decay $\sq_L^@\to q\photino$ for definiteness,
the result is the same for $\sq_R^@$ (after the exchange $L\leftrightarrow R$).

The ${\cal O}(\alpha_s)$ contribution to the $\sq_L^@ q\photino$ vertex comes
from the diagrams in Fig.~1 plus the counterterm.
Each contribution is proportional to the lowest order vertex (there is only
one Lorentz structure for the vertex because of chirality conservation).

Real gluon emission $\sq_L^@\to qg\photino$ appears at the same order and
must be added to the total rate to cancel infrared divergence.  There are
two diagrams for this process (see  Fig.~3).

The total decay rate up to ${\cal O}(\alpha_s)$ can be written as
$$\Gamma=\Gamma_0 \biggl(1+{C_F\alpha_s\over\pi}F\biggr) \;,\Eq$$
with
$$ F=F_g + F_{\gluino} + F_{\rm ren} + F_{\rm real} \;.\Eq$$
Here $F_g$, \etc, are the contributions of Fig.~1(a), (b), the counterterm,
and the real gluon emission respectively.

For clarity, we neglect the mass of the photino for a while.
The gluon-exchange (Fig.~1(a)) gives
$$F_g = {1\over2\epsilon} - \frac12\log{m_{\sq}^2\over\mu^2}
- \frac14\log^2\delta - \log\delta - {\pi^2\over4}\;.\Eq$$
Here $\delta=\lambda^2/m_{\sq}^2$,

There are two diagrams with a gluino in the loop, one with a $\sq_L$,
another with a $\sq_R$.  In the massless quark limit,
$\sq_R$ does not contribute because of chirality conservation.
It turns out that the diagram with $\sq_L$ also vanishes.  This is due to
crash between the Lorentz and chirality structure of the graph.
Hence $F_{\gluino}=0$.

The real gluon emission, integrated over the whole phase space, gives
$$F_{\rm real} = \frac14 \log^2\delta + \frac54 \log\delta + \frac{13}4
- {\pi^2\over12} \;.\Eq$$
The counterterm is defined in the same way as in the supersymmetric
case.   We calculate the quark and squark two-point functions and find
$$\eqalign{
F_{\rm ren} &= \biggl({C_F\alpha_s\over\pi}\biggr)^{-1}
\bigl[ (Z_q -1) + (Z_{\sq} -1) \bigr] \cr
&=
-{1\over2\epsilon} + \frac12\log{m_{\sq}^2\over\mu^2} - \frac14\log\delta
+ \frac34\log R - \frac12 R - 1 \cr
&\quad - \frac12 (R^2-1)\log{|R-1|\over R}
+ \frac14\biggl[ {2R-1\over(R-1)^2}\log R - {1\over R-1}\biggr] \;,
\cr}\Eqn\eqFren$$
\eqnamedef\eqtotalcorr
where $R=m_{\gluino}^2/m_{\sq}^2$.

This contribution cancels both ultraviolet and
infrared divergences, but a finite correction remains.
The total correction factor is%
\footnote{As an alternative renormalizational procedure,
we may use minimal subtraction ($\overline{\rm MS}$) to renormalize the
vertex as well as the propagators  (the counterterms just consist of poles
in $\epsilon$).  The physical $S$ matrix is then obtained by LSZ reduction
with the inclusion of the finite wave function renormalization factor.
The total result is identical with \eqtotalcorr.}
$$ F=  {3R^2-4R+2\over 4(R-1)^2} \log R -\frac12 (R^2-1)\log{|R-1|\over R}
- {2R^2-11R+10\over 4(R-1)} - {\pi^2\over3}  \;.\eqno\eqtotalcorr$$

Interestingly, \eqtotalcorr\ depends on the gluino mass
even though the loop graph with a gluino vanishes. The dependence comes
from wave function renormalization.
The mass dependence of the correction factor is shown in Fig.~4 (solid
curve).  The lowest order rate changes by about 5\%\ (for $\alpha_s\sim0.1$)
if the gluino is not too heavy.  In particular,
$F(R=0)=\frac52 - \frac{\pi^2}3 \simeq -0.790$ and
$F(R=1)=\frac{17}8 - \frac{\pi^2}3 \simeq -1.165$.
At the heavy gluino limit, we find
$$ F \simeq \frac34\log{m_{\gluino}^2\over m_{\sq}^2}
+ \frac52 - \frac{\pi^2}3 \;.\Eq$$

This logarithmic behavior can be understood in the following way.
Without supersymmetry, the photino coupling $e$ in \eqInt\ is
a Yukawa-type coupling independent of the electromagnetic gauge coupling.
If we denote the former coupling by $f$,  exact supersymmetry demands
$f=e$.  When $m_{\gluino}\gg m_{\sq}$, supersymmetry is broken at the
gluino mass scale, below which $f$ and $e$ need not be equal.
In fact, the ${\cal O}(\alpha_s)$ renormalization group equation for $f$
below $m_{\gluino}$ is found to be
$$ {df\over d\log\mu^2} = -{3C_F\over8\pi}\,\alpha_s f \;,\Eq$$
whereas the gauge coupling $e$ does not run at ${\cal O}(\alpha_s)$.
It can be seen that the logarithmic correction found in the full
calculation is nothing but the effect of the running of $f$ from the
gluino mass to the squark mass, the typical energy for the decay process.

Finally, we calculate the correction for massive photino $0<r<1$, which
may be practially important.
The diagram with a gluino is now nonzero (no divergence appears because the
amplitude is proportional to $m_{\photino}$ as well as $m_{\gluino}$).
We find
$$\eqalignno{
&F_g = {1\over2\epsilon} - \frac12\log{m_{\sq}^2\over\mu^2}
-\frac14\log^2{\delta\over(1-r)^2} - \log\delta - \Li2(r)
+ \log(1-r) -{\pi^2\over4}\;,\cr
&&\Eq\cr
&F_{\gluino} = -\sqrt{Rr} \biggl\{ {R+r-2\over(1-r)^2} H(R,r)
+{1\over r} \log(1-r) \cr
&\qquad\qquad\qquad
+ {1\over 1-r} \bigl[ R\log R - (R-1)\log|R-1|\bigr] \biggl\} \;,
&\Eq\cr
&F_{\rm real} = \frac14 \log^2{\delta\over(1-r)^2}
- \Li2(r) - \log r \log(1-r) - {\pi^2\over12} \cr
&\quad\qquad + \frac54\log\delta
-\frac52\log(1-r) - {r(4-3r)\over4(1-r)^2} \log r
+ {13-14r\over4(1-r)} \;.&\Eq\cr}$$
The function $H$ is given by
$$\eqalignno{
H(R,r) &=  \Li2({R-1\over Rr-1}) - \Li2({R+r-2\over Rr-1})
- \Li2({r(R-1)\over Rr-1}) \cr &\quad + \Li2({r(R+r-2)\over Rr-1})
\quad\hbox{(for $Rr<1$)} \cr
&=  -\Li2({Rr-1\over R-1}) + \Li2({Rr-1\over R+r-2})
+ \Li2({Rr-1\over r(R-1)}) \cr &\quad - \Li2({Rr-1\over r(R+r-2)})
- \log r \log{R+r-2\over R-1}
\quad\hbox{(for $Rr>1$)} &\Eq\cr}$$
The counterterm contribution $F_{\rm ren}$ is given by \eqFren.

%
%
The dependence of the result on $m_{\photino}$ is shown in
Figs.~4 and 5.  The logarithmic singularity at $r\to 1$ is killed by the
phase space factor in the decay rate.

\chapter{Conclusion}

We have calculated the ${\cal O}(\alpha_s)$ correction to the squark
decay width to a quark and an electroweak gaugino.  We have found that
the correction is nonzero, which can be interpreted as a manifestation
of the soft supersymmetry breaking.  In particular, logarithmic
correction appears if the gluino is heavier than the squark.  This
has an interesting implication that the supersymmetry breaking scale
may be inferred from observables at much lower energies, because
supersymmetry provides a boundary condition to connect couplings which
are otherwise unrelated.  Unfortunately, this particular example is
not very realistic in supergravity-motivated models, in which the gluino
cannot be much heavier than the squarks, with the possible exceptions of
the scalar top and bottom.

\bigskip

One of us (KH) thanks K. Higashijima, H. Murayama, Y. Okada,
K. Tobe, and T.~Yanagida for discussions.

\par \penalty-400 \vskip\chapterskip
   \spacecheck\referenceminspace \immediate\closeout\referencewrite
   \referenceopenfalse
   \leftline{\twelvebold References}\par
   \nobreak\vskip\headskip \penalty 30000
   \input reference.aux
   \endpage
\nopagenumbers
\FIG\?{One-loop diagrams for the $\bar q\sq\photino$ vertex.  The
arrow shows the flow of quark number.}
\FIG\?{One-loop diagrams for (a) quark and (b) squark self energies.}
\FIG\?{Feynman diagrams for $\sq\to q\photino g$.}
\FIG\?{$m_{\gluino}$ dependence of the correction factor $F$
for massless photino (solid), $m_{\photino}/m_{\sq}=0.2$ (dash),
0.5 (dashdot), and 0.9 (dot).}
\FIG\?{$m_{\photino}$ dependence of the correction factor $F$ for
$m_{\gluino}/m_{\sq}=0.1$ (dash), 1 (solid), 3 (dot).}
\par \penalty-400 \vskip\chapterskip
   \spacecheck\referenceminspace \immediate\closeout\figurewrite
   \figureopenfalse
   \leftline{\twelvebold Figure Captions}\par
   \nobreak\vskip\headskip \penalty 30000
   \input figures.aux

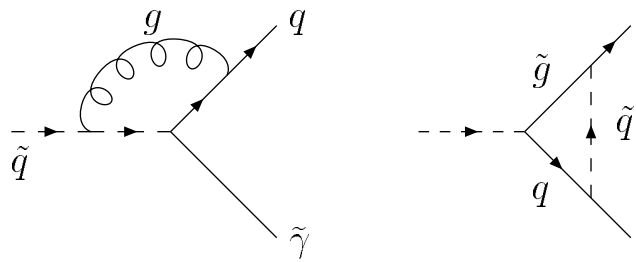
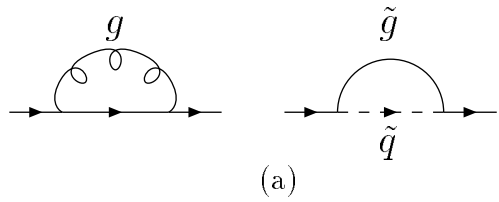
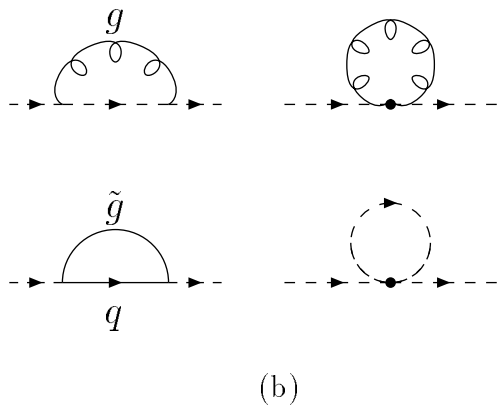
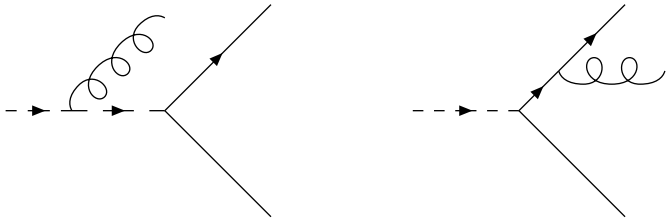
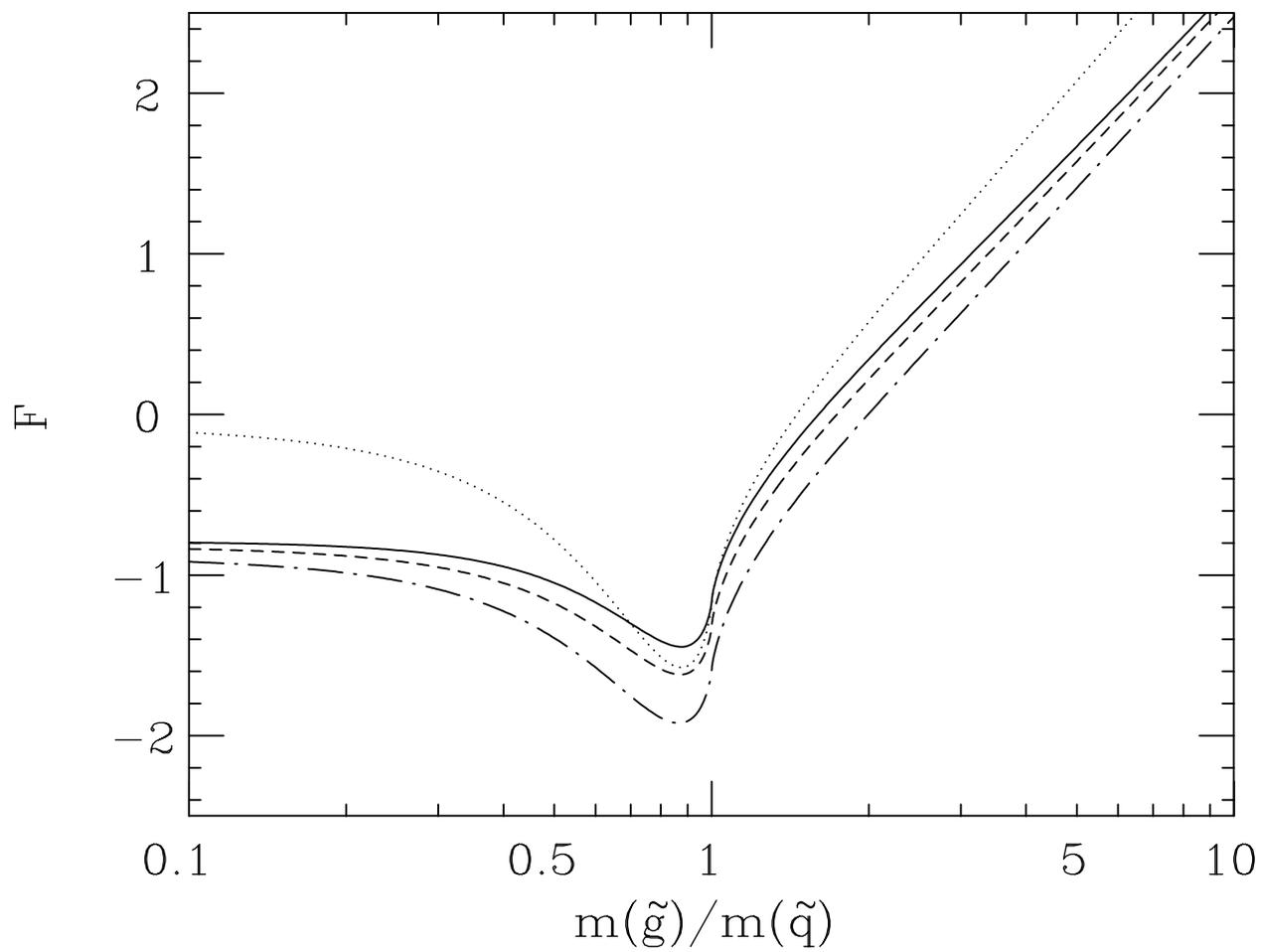
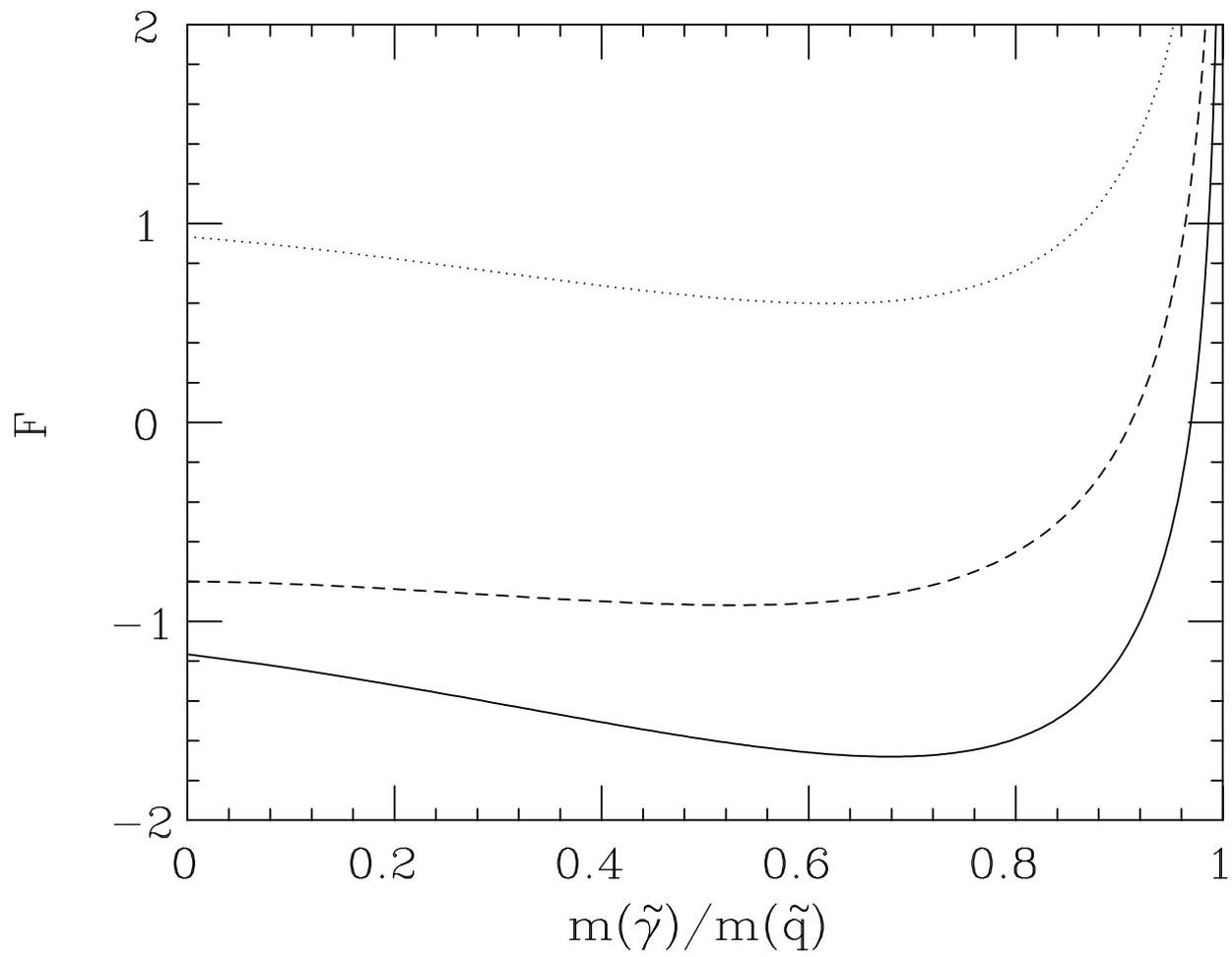
\bye